\documentclass[aps,pre,twocolumn,floatfix]{revtex4}

\usepackage{graphicx}
\usepackage{amssymb,amsfonts,amsmath}
\usepackage{epsf}
\usepackage{color}
\usepackage{subfigure}
\usepackage{epstopdf}
\usepackage{mathrsfs}

\newcommand{\be}{\begin{equation}}
\newcommand{\ee}{\end{equation}}
\newcommand{\bea}{\begin{eqnarray}}
\newcommand{\eea}{\end{eqnarray}}

\begin{document}

\title{Comment on ``Validity of path thermodynamic description of reactive systems: Microscopic simulations''}

\author{Pierre Gaspard}
\affiliation{Center for Nonlinear Phenomena and Complex Systems,\\
Universit\'e Libre de Bruxelles (U.L.B.), Code Postal 231, Campus Plaine, 
B-1050 Brussels, Belgium}

\begin{abstract}
The claims by Baras, Garcia, and Malek~Mansour [Phys. Rev. E {\bf 107}, 014106 (2023)] on the validity of path thermodynamics are ill founded and contradict well known results.  Following up on a previous comment, I show that, for both models of chemical reaction networks considered in the aforementioned paper, path thermodynamics yields values of the entropy production rates fully consistent with those expected from standard chemical thermodynamics in the large-system limit.
\end{abstract}


\maketitle

Since 2017, three papers have now been published \cite{BM17,MG20,BGM23}, where the authors reiterate the conceptual errors that reactive systems would only be described by the stochastic process of some intermediate species X and that each elementary reaction would have to be identified by the changes in the composition of this sole intermediate species X at the exclusion of the other species participating in the reaction.  As a consequence, the authors of Refs.~\cite{BM17,MG20,BGM23} overlook the consumption or production of the other species, which play essential roles in maintaining the system away from equilibrium, thereby causing the production of entropy.

In my previous Comment~\cite{G21}, I explained that, according to path thermodynamics, it is necessary to account for all species, whether intermediate or not, in order to identify the sequence of elementary reactions taking place so as to evaluate the rate of entropy production, as formulated, in particular, in Ref.~\cite{G04}.  In view of their recent publication~\cite{BGM23}, the authors gave no consideration to my refutation of their faulty argument against the validity of path thermodynamics.  In this Comment, I summarize the claims of Ref.~\cite{BGM23} and show that it contains contradictions with the Schnakenberg 50 years old theory of Markov jump processes~\cite{S76}.  Moreover, I also show that, for the models considered in Ref.~\cite{BGM23}, the values of entropy production computed within the formalism of path thermodynamics are fully consistent with the standard values predicted by chemical thermodynamics, provided its principles are duly respected.

In short, numerical results are presented in Ref.~\cite{BGM23} for two separate models of chemical reaction networks (see Table~\ref{table}), based on Bird's Direct Simulation Monte Carlo algorithm.  Their point is to test the reversibility for the stochastic process $\{X(t)\}$ followed by the intermediate species X.  The two models are chosen because, in model~I, the intermediate species X has the same stoichiometric coefficient ($\nu_{{\rm X},+1}=\nu_{{\rm X},+2}=1$) in both forward reactions, whereas, in model~II, they are different ($\nu_{{\rm X},+1}=2$, $\nu_{{\rm X},+2}=1$).  The results reported in Ref.~\cite{BGM23} show that the stochastic process $\{X(t)\}$ of the intermediate species X is reversible in model~I, but not in model~II, even though both are out of equilibrium.  The authors wrongly conclude that these facts invalidate path thermodynamics.

Before I proceed, I should emphasize that the reversibility of stochastic processes generated by some random variables is by no means in contradiction with the process being out of equilibrium.  As a matter of fact,  I~gave two examples of such processes in my previous Comment~\cite{G21}, where I~mentioned that the stochastic process $\{X(t)\}$ of the intermediate species X in the simple reaction network ${\rm A} \rightleftharpoons {\rm X} \rightleftharpoons {\rm B}$ is indeed reversible even when the system is subject to nonequilibrium conditions.  In the same way, the velocity $\{v(t)\}$ of a Brownian particle driven away from equilibrium by a constant external force can follow a reversible process although the joint variables of position and velocity $\{r(t),v(t)\}$ do not.  

Nobody denies the validity of these facts, which have helped formulate path thermodynamics for reactive systems.  Rather, they show the limitations of restricting the description of reactive systems to the sole intermediate species X and the need, in chemical thermodynamics, to consider all the species in order to identify each elementary reaction \cite{KP98,JVN84}.  The reason is that the different elementary reactions are distinguished by the changes in composition of all the chemical species involved in each reaction. For models I and II described in Table~\ref{table}, these species are (X,A,B,C), which may undergo the changes $(X,A,B,C)\overset{\rho}\longrightarrow (X+\nu_{{\rm X}\rho}, A + \nu_{{\rm A}\rho}, B + \nu_{{\rm B}\rho}, C+\nu_{{\rm C}\rho})$, where $(\nu_{{\rm X}\rho}, \nu_{{\rm A}\rho}, \nu_{{\rm B}\rho}, \nu_{{\rm C}\rho})$ are the stoichiometric coefficients of the species in the reaction~$\rho$.  Even if the species (A,B,C) are kept at constant concentrations, the reactions  can continuously consume (or produce) molecules in the pool of these species, so that the system should be open to inlet and outlet mass flows.  Therefore, at the level of successive reactions, the numbers $(A,B,C)$, which count the molecules with respect to some initial values, change in time and we should {\it a priori} consider the complete stochastic process formed by the joint trajectories $\{X(t),A(t),B(t),C(t)\}$.

\begin{table*}[ht]
\caption{\label{table} Kinetics and thermodynamics for the two models compared in Ref.~\cite{BGM23}. The reaction constants are denoted $k_{\rho}$ with $\rho=\pm1,\pm2$, $\Omega$ is the extensivity parameter, $X$ is the number of molecules of the intermediate species, and $a$, $b$, $c$ are the molecule fractions of the chemostatted species.  ${\cal A_C}$ and ${\cal J_C}$ are respectively the affinity and the overall rate associated with the given cycle ${\cal C}$.  Similar results hold for other cycles. For the intermediate species X, the stationary value of the molecule fraction is given by $x_s=\langle X\rangle_s/\Omega$.}
\vspace{5mm}
\begin{center}
\begin{tabular}{|c|c|c|}
\hline
 & Model I & Model II \\
\hline
         &      &     \\
reaction network & ${\rm A} + {\rm X} \, \underset{k_{-1}}{\overset{k_{+1}}{\rightleftharpoons}}\, 2 \, {\rm X}$ & $2 \, {\rm A} \, \underset{k_{-1}}{\overset{k_{+1}}{\rightleftharpoons}}\, 2 \, {\rm X}$ \\
 & ${\rm B} + {\rm C} \, \underset{k_{-2}}{\overset{k_{+2}}{\rightleftharpoons}}\, {\rm B}+ {\rm X}$ & ${\rm B} + {\rm C} \, \underset{k_{-2}}{\overset{k_{+2}}{\rightleftharpoons}}\, {\rm B}+ {\rm X} $\\
         &      &     \\
rates & $W_{+1}(X) = k_{+1} \, a\,  X$ & $W_{+1}(X) = k_{+1} \, a^2 \, \Omega$\\
         & $W_{-1}(X) = k_{-1} X (X-1)/\Omega$ \ & \ $W_{-1}(X) = k_{-1} X (X-1)/\Omega$ \\
         & $W_{+2}(X) = k_{+2}\, b\,  c \, \Omega$ & $W_{+2}(X) = k_{+2}\, b \, c \, \Omega$ \\
         & $W_{-2}(X) = k_{-2}\, b\, X$ & $W_{-2}(X) = k_{-2}\, b \, X$ \\
         &      &     \\
cycle & ${\cal C}= X  \, {\overset{\rho=+1}{\longrightarrow}}\, X+1  \, {\overset{\rho=-2}{\longrightarrow}}\, X$ & ${\cal C}= X  \, {\overset{\rho=+1}{\longrightarrow}}\, X+2  \, {\overset{\rho=-2}{\longrightarrow}}\, X+1 \, {\overset{\rho=-2}{\longrightarrow}}\, X$ \\
         &      &     \\
cycle affinity & $\displaystyle {\cal A_C} = \ln\frac{k_{+1} k_{-2}\, a}{k_{-1} k_{+2}\, c}$ & $\displaystyle {\cal A_C} = \ln\frac{k_{+1} k_{-2}^2 \, a^2}{k_{-1} k_{+2}^2 \, c^2}$ \\
         &      &     \\
overall rate &\ ${\cal J_C} = k_{+1} a x_s - k_{-1} x_s^2 = -(k_{+2} bc - k_{-2} b x_s)$ \ & \ ${\cal J_C} = k_{+1} a^2 - k_{-1} x_s^2 = -(k_{+2} bc - k_{-2} b x_s)/2$ \ \\
         &      &     \\
in/out flows & $(d/dt)\langle A\rangle_s = - (d/dt)\langle C\rangle_s = - \Omega \, {\cal J_C} $ & $(d/dt)\langle A\rangle_s = - (d/dt)\langle C\rangle_s = - 2\, \Omega \, {\cal J_C} $ \\
         &      &     \\
\hline
\end{tabular}
\end{center}
\end{table*}

The reactive systems considered in Ref.~\cite{BGM23} may be described by Markov jump processes with the transition rates given in Table~\ref{table} for both models.  The central issue however is that the complete stochastic process for the joint probability distribution $P(X,A,B,C,t) \equiv P(X,\pmb{n},t)$ with $\pmb{n}=(A,B,C)$ is ruled by the master equation
\bea
&& \frac{d}{dt} P(X,\pmb{n},t) \nonumber\\
&& = \sum_{\rho=\pm 1}^{\pm r} \Big[ W_{\rho}(X-\nu_{{\rm X}\rho})\; 
P(X-\nu_{{\rm X}\rho},\pmb{n} -\Delta\pmb{n}_{\rho},t) \nonumber\\
&&\qquad\quad - W_{-\rho}(X)\; P(X,\pmb{n},t) \Big]
\qquad
\label{full.master.eq}
\eea
with $r=2$ and the stoichiometric coefficients $\Delta\pmb{n}_{\rho}=(\nu_{{\rm A}\rho}, \nu_{{\rm B}\rho}, \nu_{{\rm C}\rho})$ and not by the reduced master equation
\bea
\frac{d}{dt} P(X,t) &=& \sum_{\rho=\pm 1}^{\pm r} \Big[ W_{\rho}(X-\nu_{{\rm X}\rho})\; 
P(X-\nu_{{\rm X}\rho},t) \nonumber\\
&&\qquad - W_{-\rho}(X)\; P(X,t)\Big] ,
\qquad
\label{master.eq}
\eea
for the marginal probability distribution of the process $\{X(t)\}$, $P(X,t) = \sum_{A,B,C} P(X,A,B,C,t) \equiv \sum_{\pmb{n}} P(X,\pmb{n},t)$.
Equation~(\ref{full.master.eq}) is equivalent to Eq.~(19) of my previous Comment~\cite{G21}, and is well known to be the basic master equation of the complete process \cite{McQ67,NP77,vK83,H83,Gardiner}.  Contrary to what is assumed in Refs.~\cite{BM17,MG20,BGM23}, the reduced master equation~(\ref{master.eq}) is not the unique master equation for the description of the reactive system.  There is no contradiction with any fundamental property of jump processes in considering  the complete master equation~(\ref{full.master.eq}) alongside the reduced master equation~(\ref{master.eq}), which refer to processes with different state spaces.

In the case of model~I, the reduced master equation can be directly expressed in terms of the rates $W_{\pm}\equiv W_{\pm 1}+W_{\pm 2}$ for the transitions $X\longrightarrow X\pm 1$, which actually defines a new model, herein called model~$0$, where the two elementary reactions $1$ and $2$ are no longer distinguishable.  In the case of model~II, the two elementary reactions remain distinguishable and they cannot be lumped together, because they correspond to the different transitions $X\longrightarrow X\pm 2$ and $X\longrightarrow X\pm 1$, respectively.

However, for the sake of path thermodynamics, we must always identify the elementary reactions involved in the random jumps of the process in order to evaluate the entropy produced during the time evolution.  That is to say, we should consider the joint stochastic trajectories $\{X(t),A(t),B(t),C(t)\}$, as mentioned earlier.  The knowledge of these trajectories is equivalent to knowing the joint sequence $\{X_l,\rho_l\}$ of the numbers $X_l$ of molecules between the jumps and the successive elementary reactions $\rho_l$ causing these jumps.  The entropy production rate can thus be computed using the stochastic formulation of path thermodynamics by Lebowitz and Spohn \cite{LS99}.  In this formulation, the entropy production rate is given by \cite{G04}
\be
\frac{1}{k_{\rm B}} \frac{d_{\rm i}S}{dt} = \lim_{t\to\infty} \frac{1}{t} \sum_{l=1}^{n(t)} \ln\frac{W_{\rho_l}(X_l)}{W_{-\rho_l}(X_l+\nu_{{\rm X}\rho_l})}\, ,
\label{EPR-stoch}
\ee
where $W_{\rho_l}(X_l)$ denotes the transition rate of the elementary reaction $X_l \overset{\rho_l} \longrightarrow X_l+\nu_{{\rm X}\rho_l}$ and $W_{-\rho_l}(X_l+\nu_{{\rm X}\rho_l})$ the rate of the reversed reaction $X_l+\nu_{{\rm X}\rho_l}\overset{-\rho_l}\longrightarrow X_l$, $k_{\rm B}$~is~Boltzmann's constant, and the sum is carried out over the sequence of $n(t)$ elementary reactions $\{\rho_l\}$ occurring in time $t$.  In the large-system limit $\Omega\gg 1$, Eq.~(\ref{EPR-stoch}) becomes equal to the standard formula of chemical thermodynamics for the entropy production rate,
\be
\frac{1}{k_{\rm B}} \frac{d_{\rm i}S}{dt} = \Omega \sum_{\rho=1}^{r} \left( w_{+\rho}-w_{-\rho}\right) \ln\frac{w_{+\rho}}{w_{-\rho}} \, ,
\label{EPR-thermo}
\ee
where the reaction rates are defined in terms of the transition rates, according to $w_{\rho}(x)\equiv\lim_{\Omega\to\infty}\Omega^{-1} W_{\rho}(\Omega x)$.  It is straightforward to deduce the result~(\ref{EPR-thermo}) from Eq.~(\ref{EPR-stoch}), since $\Omega( w_{+\rho}-w_{-\rho})$ is the net occurrence rate of $\ln(W_{+\rho}/W_{-\rho}) \simeq\ln(w_{+\rho}/w_{-\rho})$ in the sum of Eq.~(\ref{EPR-stoch}) for every elementary reaction $1\le \rho\le r$ \cite{G04}.  If, in contrast, the sum in Eq.~(\ref{EPR-stoch}) were restricted to the sole transitions causing the changes $X\longrightarrow X+\Delta X$ with integer values $\Delta X$ for the intermediate species X, the entropy production rate~(\ref{EPR-thermo}) would not be given by the sum of the entropies produced by each elementary reaction, which would be in contradiction with standard chemical thermodynamics \cite{KP98}.  By refusing to consider the former approach and arguing that only the latter is possible, the authors of Refs.~\cite{MG20,BGM23} are bound to reach inconsistent conclusions.

Let us further remark that several Markov jump processes may be considered for a given reaction network.  This key point is well known.  According to Schnakenberg's theory \cite{S76}, a graph can be associated with a Markov jump process by assigning vertices to each of the states and edges to the allowed transitions between the states.  The graph associated with the model formed by the reactions ${\rm A}+ 2 {\rm X} \rightleftharpoons 3{\rm X}$ and ${\rm B}+{\rm C} \rightleftharpoons {\rm B}+{\rm X}$ (which is similar to model~I) is shown in Fig.~2 of the paper~\cite{S76} and this graph presents two edges connecting every pair of vertices.  As a consequence, there exist cycles in such graphs, which allow us to define the affinities driving the system out of equilibrium.  In a steady state, the entropy production rate (\ref{EPR-thermo}) can be expressed as the product of the affinity ${\cal A_C}$ and the mean overall rate ${\cal J_C}$ associated with some cycle ${\cal C}$ according to
\be
\frac{1}{k_{\rm B}} \frac{d_{\rm i}S}{dt} = \Omega \, {\cal A_C} \, {\cal J_C} \, .
\label{EPR-stst}
\ee
See Table~\ref{table} for application to models I and II.
Thus, Schnakenberg showed in 1976 that a Markov jump process may be defined in such a way that the two elementary reactions can always be distinguished, which invalidates what is asserted in Ref.~\cite{BGM23}.

\begin{figure}[htbp]
\centerline{\includegraphics[width=10cm]{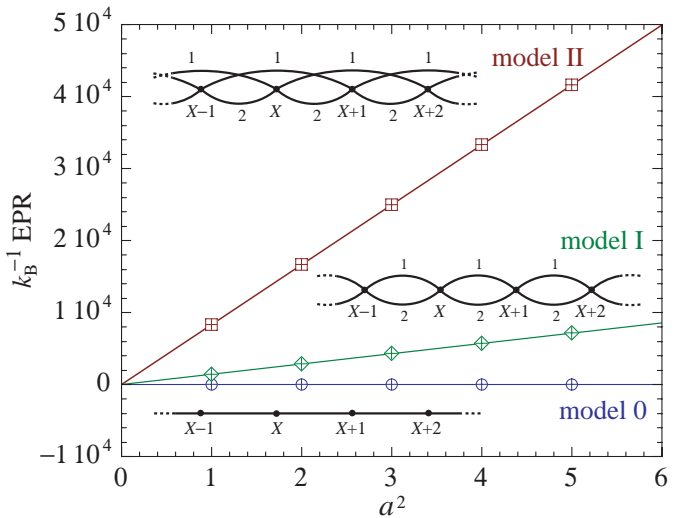}}
\caption{Entropy production rate (EPR $=d_{\rm i}S/dt$) versus $a^2$ for model~$0$ (open circles), model~I (open diamonds), and model~II (open squares) computed with the stochastic method using Eq.~(\ref{EPR-stoch}).  The pluses joined by the solid lines give the expectation (\ref{EPR-thermo}) from standard macroscopic thermodynamics.  The extensivity parameter is equal to $\Omega=10^4$ and the total time interval to compute the EPR is taken as $t=10^4$.  For model~I, the parameter values are $k_{+1}=k_{-1}=1$, $k_{+2}=k_{-2}=5/6$, $b=6a/5$, $c=a/2$, the steady state is $x_s=a/\sqrt{2}$, and the expected entropy production rate (\ref{EPR-thermo}) is given by $k_{\rm B}^{-1}{\rm EPR}_{\rm th}=(\Omega/2)a^2(\sqrt{2}-1)\ln 2\simeq 0.14356\, a^2 \Omega$.  For model~II, they are $k_{+1}=k_{-1}=k_{+2}=k_{-2}=1$, $a=1$, $b=5a/3$, $c=a/3$, $x_s=a(\sqrt{209}-5)/12$, and $k_{\rm B}^{-1}{\rm EPR}_{\rm th}=\Omega(a^2-x_s^2)\ln 9\simeq 0.83263\, a^2 \Omega$. For model~$0$, the parameter values and the steady state are the same as for model~I, but ${\rm EPR}_{\rm th}=0$.  The mean accuracy $k_{\rm B}^{-1}\langle\vert{\rm EPR}_{\rm num}-{\rm EPR}_{\rm th}\vert\rangle$ is equal to $3\times 10^{-6}$, $0.75$, and $2.6$ for models $0$, I, and II, respectively.  The Hill-Schnakenberg graphs of the models are shown as insets.}
\label{fig1}
\end{figure}

Figure~\ref{fig1} shows the values of the entropy production rates computed using~(\ref{EPR-stoch}) and~(\ref{EPR-thermo}) for the models $0$, I, and~II, with the corresponding Schnakenberg graphs displayed as insets.  On top of the values predicted by Eq.~(\ref{EPR-thermo}), we report the numerical results of simulations of the stochastic processes here performed with Gillespie's algorithm, which is more readily applicable and much faster than Bird's algorithm, allowing larger molecule numbers on a personal computer.  The entropy production rate of path thermodynamics is computed with Eq.~(\ref{EPR-stoch}) and plotted as the open symbols in Fig.~\ref{fig1}.  We observe the excellent agreement with the expectations from standard thermodynamics (pluses joined by solid lines) as given by Eq.~(\ref{EPR-thermo}) or (\ref{EPR-stst}).  Away from equilibrium, the affinity ${\cal A_C}$ and the overall rate ${\cal J_C}$ are different from zero and the entropy production rate is positive in both models~I and~II.  On the contrary, the graph of model~$0$ has no cycle, so that no affinity can be defined for it, which therefore behaves as an equilibrium process. 

These results show that path thermodynamics is perfectly valid for both models~I and~II, which have the expected positive entropy production rate under nonequilibrium conditions.  However, the entropy production rate of model~$0$, i.e., the lumped model~I as considered in Ref.~\cite{BGM23}, is equal to zero because that jump process does not make the distinction between the two elementary reactions of model~I.  

Furthermore and quite disappointingly, there is no attempt in the numerical results presented in Ref.~\cite{BGM23} to evaluate the entropy production rate of models I and II, and to test proposals that have been published in the literature.  As a matter of fact, the computation of entropy production can be performed using any simulation algorithm for the reason that the transition rates of the elementary reactions are necessarily defined within the algorithm, so that the inlet and outlet mass flows of the chemostatted species can also be measured.  Indeed, reaction networks are driven away from equilibrium by the mass flows required to chemostat some species (i.e., A, B, and C in this case).  Although the fractions of chemostatted species are kept invariant in time, these species may be consumed or produced at non-zero rates, unless the system is in equilibrium and the detailed balance conditions hold.  In models I and~II, the species B is spectator for the two reactions, so that $B(t)$ remains constant and $(d/dt)\langle B\rangle=0$.  However, the stochastic processes $\{A(t)\}$ and $\{C(t)\}$ are non stationary if the reactive system is driven away from equilibrium.  In so-called steady states, the stationary condition for the intermediate species X is satisfied because $(d/dt)\langle X\rangle_s=-(d/dt)\langle A\rangle_s-(d/dt)\langle C\rangle_s=0$.   The consumption or production rates of the chemostatted species A and C are given in Table~\ref{table} for models I and~II in the limit $\Omega\gg 1$.  In this regard, the species A and C act as fuel and product sustaining the nonequilibrium conditions.

To conclude, the criticisms expressed in Ref.~\cite{BGM23} about the validity of path thermodynamics are unfounded. The fundamental principles of chemical thermodynamics should not be ignored.  The issue is very concrete and concerns tangible quantities in the real world, as shown by many examples of everyday life.  For instance, in electric circuits such as a resistor, a diode, or an electrolytic cell connected to a battery, the entropy production rate is given by $d_{\rm i}S/dt=P/T=VI/T$ in terms of the dissipated power $P=VI$, the temperature $T$, the applied voltage $V$, and the electric current~$I$ across the device.  In this analogy, the electric current $I$ corresponds to $e(d/dt)\langle C\rangle$ and the affinity is given by ${\cal A_C}=eV/(k_{\rm B}T)$, where $e$ denotes the elementary electric charge.  The concentration of ions may significantly differ depending on the type of cells, but the entropy production rate is always determined by the applied voltage and the corresponding electric current.  Another example is home heating.  Thermal insulation can have different efficiencies and the outside temperature may vary with the seasons.  Yet, the thermodynamic cost of heating is better assessed by the fuel gauge than the internal temperature, i.e., by $\vert\langle C\rangle_t-\langle C\rangle_0\vert$ rather than $\langle X\rangle$ in this other analogy.  Really, the issue is of concern to all of us.


\section*{Acknowledgments}

The author thanks Thomas Gilbert for valuable suggestions.
This research is supported by the Universit\'e Libre de Bruxelles (ULB).


\end{document}